\documentclass[12pt]{article}
\usepackage{epsf}
\title{ {\bf
Anomalous magnetic moment of the muon in the two Higgs doublet 
model}}
\author{\vspace{1cm}\\
        {\bf E. O. Iltan}
        \thanks{E-mail address:
        eiltan@heraklit.physics.metu.edu.tr} 
\\
Physics Department, Middle East Technical University \\
        Ankara, Turkey \\ \\
        {\bf H. Sundu}
        \thanks{E-mail address:
        sundu@metu.edu.tr}
 \\
        Physics Department, Middle East Technical University \\
        Ankara, Turkey\\}

\date{}

\begin{document}
\setlength{\baselineskip}{24pt}
\maketitle
\setlength{\baselineskip}{7mm}
\begin{abstract}
We calculate the new physics effects on the anomalous magnetic moment of 
the muon in the framework of the two Higgs doublet model. We predict 
an upper bound for the lepton flavor violating coupling, which is
responsible for the point like interaction between muon and tau, by using 
the uncertainty in the experimental result of the muon anomalous magnetic
moment. We show that the upper bound predicted is more stringent compared to
the one which is obtained by using the experimental result of the muon 
electric dipole moment. 
\end{abstract} 
\thispagestyle{empty}
\newpage
\setcounter{page}{1}
\section{Introduction}
The lepton flavor violating (LFV) interactions, non-zero electric dipole moments 
(EDM) and the anomalous magnetic moments (AMM) of leptons  are among the most 
promising candidates to search for physics beyond the standard model (SM). 
The AMM of the muon have been studied in the literature extensively \cite{Nie} and 
\cite{SUSY}. The experimental result of the muon AMM by the g-2 Collaboration 
\cite{Brown} has been obtained as
\begin{eqnarray}
a_{\mu}=116\, 592\, 023\, (151)\times 10^{-11}\,\, ,
\end{eqnarray}
and recently, at BNL \cite{BNL}, a new experimental world average 
has been announced 
\begin{eqnarray}
a_{\mu}=11\, 659\, 203\, (8)\times 10^{-10}\,\, , 
\end{eqnarray}
which has about half of the uncertainty of previous measurements.
This result has opened a new window for testing the SM and the new physics 
effects beyond. The SM prediction for $a_{\mu}$ is written in terms of
different contributions \cite{Davier}; 
\begin{eqnarray}
a_{\mu}(SM)=a_{\mu}(QED)+a_{\mu}(weak)+a_{\mu}(hadronic)
\,\, , 
\end{eqnarray}
where $a_{\mu}(QED)=11\, 658\,470.57\, (0.29)\times 10^{-10}$ and 
$a_{\mu}(weak)=15.1\,(0.4)\times 10^{-10}$. The hadronic contributions are
under theoretical investigation. With the new data from Novosibirsk
\cite{Akhmetshin}, the calculation of the first order hadronic vacuum
polarization to $a_{\mu}(SM)$ is obtained as 
$684.7\,(7.0)\times 10^{-10}$ ($701.9\,(6.1)\times 10^{-10}$) using the 
$e^+ e^- $ $(\tau)$ based result. The addition of the higher order
contributions, $-10.0\,(0.6)\times 10^{-10}$ and light by light scattering   
$-8.6\,(3.2)\times 10^{-10}$,   result in  
$a_{\mu}(SM)=11\, 659\, 169.1\, (7.8)\times 10^{-10}$ 
($a_{\mu}(SM)=11\, 659\, 186.3\, (7.1)\times 10^{-10}$ based on  
$e^+ e^-$ $(\tau)$ data. Therefore, there is a $3.0$ ($1.6$) standard 
deviation from the experimental result and this could possibly be due 
to the effects of new physics, at present.   

Various scenarios have been proposed to explain the nonvanishing value of
the deviation $\Delta a_{\mu}$ \cite{Lane} - \cite{Pires},previously. The 
Supersymmetry (SUSY) contribution to $a_{\mu}$ has been investigated in 
\cite{SUSY,Feng, Komine}. In \cite{Gninenko} the new physics effect on 
$a_{\mu}$ has been explained by introducing a new light gauge boson. The 
prediction of the muon AMM has been estimated in the framework of leptoquark 
models in \cite{Cheung}, the technicolor model with scalars and top color 
assisted technicolor model in \cite{Xiong}, in the  framework of the general 
two Higgs doublet model (2HDM) in \cite{Dede} and also in \cite{Diaz}. The 
work \cite{Kang} was devoted to the Higgs mediated lepton flavor violating 
interactions which contributed to $a_{\mu}$. In this study, only the scalar 
Higgs exchange was taken into account by assuming that the pseudoscalar Higgs 
particle was sufficiently heavier than the scalar one. Finally, in 
\cite{Pires}, scalar scenarios contributing to $a_{\mu}$ with enhanced Yukawa 
coupling were proposed. 

In \cite{Nie}, the upper bound on leptonic flavor changing coupling, related 
with the transition $\tau-\mu$, has been obtained in the 2HDM as $0.11$, using 
the AMM of the muon by considering that the dominant contribution comes from 
the lighter scalar boson. In this case the uncertainty between the SM
prediction and the experimentalone was taken as $7.4\times 10^{-9}$ and it was 
emphasized that this bound would decrease to the values of $\sim 0.03$ with 
the reduction of the uncertainty up to a factor $20$.  

In our work, we study the new physics effects on the AMM of the muon using 
the model III version of the 2HDM of reference \cite{Nie}, including both 
scalar and pseudoscalar Higgs boson effects, based on the assumption that 
the numerical value should not exceed the present experimental uncertainty, 
$\sim (1-2) \times 10^{-9}$. The new contribution to $a_{\mu}$ exists at 
one-loop level with internal mediating neutral particles $h^0$ and $A^0$ 
in our case, since we do not include charged FC interaction in the leptonic 
sector due to the small couplings for $\mu-\nu_l$ interactions. In the 
calculations, we take into account the internal $\tau$ and $\mu$ leptons 
and neglect the contribution coming from the internal $e$-lepton since the 
corresponding Yukawa coupling is expected to be smaller compared to the 
others. Furthermore, we also neglect the internal $\mu$-lepton contribution 
by observing the weak dependence of $\Delta a_{\mu}$ on the $\mu$-$\mu$ 
coupling. We predict a stringent upper bound for the $\mu$-$\tau$ coupling 
and compare with the one, which is obtained by using the restriction coming 
from the EDM of $\mu$ lepton (see \cite{Iltan} for details). 

The paper is organized as follows: In Section 2, we present the new physics
effects on the AMM of the muon in the framework of the general 2HDM. Section 
3 is devoted to discussion and our conclusions.
\section{Anomalous magnetic moment of the muon in the model III version of 
two Higgs doublet model.} 
In the type III 2HDM, there exist flavor changing neutral currents 
(FCNC), mediated by the new Higgs bosons, at tree level. The most general
Higgs-fermion interaction for the leptonic sector in this model reads as 
\begin{eqnarray}
{\cal{L}}_{Y}=
\eta^{E}_{ij} \bar{l}_{i L} \phi_{1} E_{j R}+
\xi^{E}_{ij} \bar{l}_{i L} \phi_{2} E_{j R} + h.c. \,\,\, ,
\label{lagrangian}
\end{eqnarray}
where $i,j$ are family indices of leptons, $L$ and $R$ denote chiral 
projections $L(R)=1/2(1\mp \gamma_5)$, $l_{i L}$ and $E_{j R}$ are 
lepton doublets and singlets respectively, $\phi_{i}$ for $i=1,2$, are the 
two scalar doublets 
\begin{eqnarray}
\phi_{1}=\frac{1}{\sqrt{2}}\left[\left(\begin{array}{c c} 
0\\v+H^{0}\end{array}\right)\; + \left(\begin{array}{c c} 
\sqrt{2} \chi^{+}\\ i \chi^{0}\end{array}\right) \right]\, ; 
\phi_{2}=\frac{1}{\sqrt{2}}\left(\begin{array}{c c} 
\sqrt{2} H^{+}\\ H_1+i H_2 \end{array}\right) \,\, ,
\label{choice}
\end{eqnarray}
with the vacuum expectation values   
\begin{eqnarray}
<\phi_{1}>=\frac{1}{\sqrt{2}}\left(\begin{array}{c c} 
0\\v\end{array}\right) \,  \, ; 
<\phi_{2}>=0 \,\, .
\label{choice2}
\end{eqnarray}
With the help of this parametrization and 
considering the gauge and $CP$ invariant Higgs potential which 
spontaneously breaks  $SU(2)\times U(1)$ down to $U(1)$  as:
\begin{eqnarray}
V(\phi_1, \phi_2 )&=&c_1 (\phi_1^+ \phi_1-v^2/2)^2+
c_2 (\phi_2^+ \phi_2)^2 \nonumber \\ &+& +
c_3 [(\phi_1^+ \phi_1-v^2/2)+ \phi_2^+ \phi_2]^2
+ c_4 [(\phi_1^+ \phi_1) (\phi_2^+ \phi_2)-(\phi_1^+ \phi_2)(\phi_2^+ \phi_1)]
\nonumber \\ &+& 
c_5 [Re(\phi_1^+ \phi_2)]^2 +
c_{6} [Im(\phi_1^+ \phi_2)]^2 
+c_{7}\,\, ,
\label{potential}
\end{eqnarray}
the SM particles and new particles beyond can be collected in the first
and 
second doublets respectively. Here $H^0$ is the SM Higgs boson and 
$H_1\,(H_2)$ are the new neutral Higgs particles. Since there is no mixing of 
neutral Higgs bosons at tree level for this choice of Higgs doublets, 
$H_1\,(H_2)$  is the usual scalar (pseudoscalar) $h^0\,(A^0)$. 

In the Yukawa interaction eq. (\ref{lagrangian}), the part which is responsible 
for the FCNC at tree level reads as
\begin{eqnarray}
{\cal{L}}_{Y,FC}=
\xi^{E}_{ij} \bar{l}_{i L} \phi_{2} E_{j R} + h.c. \,\, .
\label{lagrangianFC}
\end{eqnarray}
Notice that, in the following we will replace $\xi^{E}_{ij}$ by 
$\xi^{E}_{N,ij}$ to emphasize that the couplings are related to the 
neutral interactions. The Yukawa matrices $\xi^{E}_{N,ij}$ have in 
general complex entries and they are free parameters which should be fixed 
by using the various experimental results. 

The effective interaction for the anomalous magnetic moment of the lepton 
is defined as 
\begin{eqnarray}
{\cal L}_{AMM}=a_l \frac{e}{4\,m_l}\,\bar{l} \,\sigma_{\mu\nu}\,l\, F^{\mu\nu} 
\,\, ,
\label{AMM1}  
\end{eqnarray}
where $F_{\mu\nu}$ is the electromagnetic field tensor and "$a_l$" is the AMM 
of the lepton "$l$", $(l=e,\,\mu,\,\tau)$. This interaction can be induced by 
the neutral Higgs bosons $h^0$ and $A^0$ at loop level in the model III, 
beyond the SM. As mentioned we do not take charged FC interaction in the 
leptonic sector due to the small couplings for $\mu-\nu_l$ interactions.

\begin{figure}[htb]
\vskip -0.0truein
\centering
\epsfxsize=6.0in
\leavevmode\epsffile{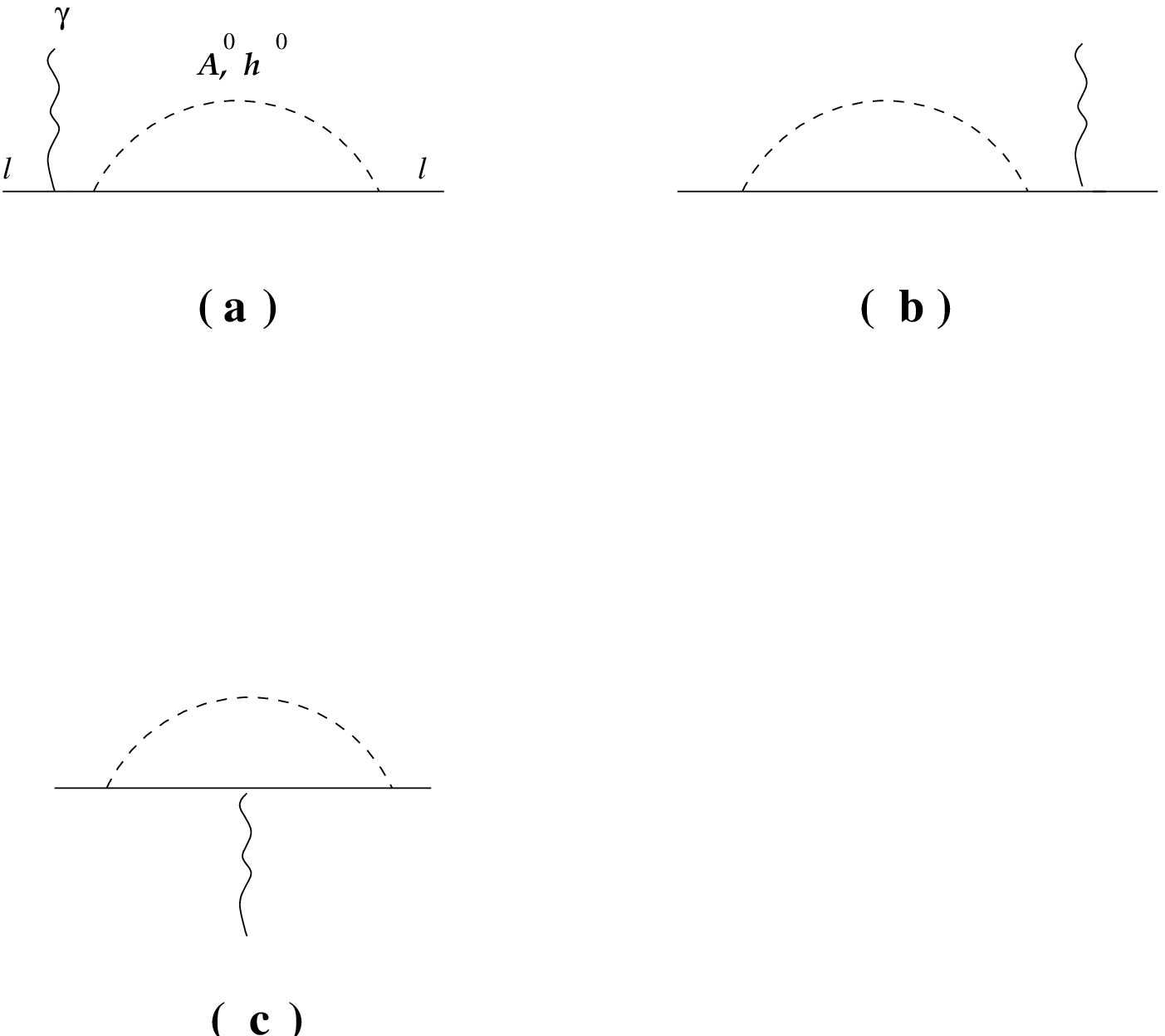}
\vskip 0.5truein
\caption[]{One loop diagrams contributing to AMM of $l$-lepton  due to the 
neutral Higgs bosons $h^0$ and $A^0$ in the 2HDM. Wavy (dashed) line 
represents the electromagnetic field ($h^0$ or $A^0$ fields).}
\label{fig1}
\end{figure}

In Fig. \ref{fig1}, we present the 1-loop diagrams due to neutral Higgs 
particles .  Since, the self energy $\sum(p)$ (diagrams $a$, $b$ in Fig. 
\ref{fig1}) vanishes when the $l$-lepton is on-shell, in the on-shell 
renormalization scheme, only the vertex diagram $c$ in Fig. \ref{fig1}
contributes to the calculation of the AMM of the lepton $l$. The most general 
Lorentz-invariant form of the coupling of a charged lepton to a photon of 
four-momentum $q_{\nu}$ can be written as
\begin{eqnarray}
\Gamma_{\mu}&=& G_1 (q^2)\, \gamma_{\mu} + G_2 (q^2)\, \sigma_{\mu\nu} 
\,q^{\nu} 
\nonumber \\ &+& 
G_3 (q^2)\, \sigma_{\mu\nu}\gamma_5\, q^{\nu}
\label{vertexop}
\end{eqnarray}
where $q_{\nu}$ is the photon 4-vector and the $q^2$ dependent form factors 
$G_{1}(q^2)$, $G_{2}(q^2)$ and $G_{3}(q^2)$ are proportional to the charge, 
AMM and EDM of the $l$-lepton respectively. Using the definition of AMM of the 
lepton $l$ (eq. (\ref{AMM1})), $\Delta_{New} a_{\mu}$  is extracted as
\begin{eqnarray}
\Delta_{New} a_{\mu}=a_{\mu}^{(1)}+\int_0^1\, a_{\mu}^{(2)}(x)\, dx \,\, ,
\label{tauANOM}
\end{eqnarray}
where $a_{\mu}^{(1)}$ ($\int_0^1\, a_{\mu}^{(2)}(x)\, dx$) is the 
contribution coming from the internal $\tau$ ($\mu$) lepton. The functions 
$a_{\mu}^{(1)}$ and $a_{\mu}^{(2)}$ are given by
\begin{eqnarray}
a_{\mu}^{(1)} &=& \frac{G_F}{\sqrt{2}}\, \frac{Q_{\tau}}{64\, \pi^2}\, 
\Big{\{} \frac{1}{2} ((\bar{\xi}^{E\,*}_{N,\mu\tau})^2 + 
(\bar{\xi}^{E}_{N,\tau \mu})^2)\, (F_1 (y_{h^0})-F_1 (y_{A^0}))
\nonumber \\ &+& 
\frac{1}{3} |\bar{\xi}^{E}_{N,\tau\mu}|^2\,
\frac{m_{\mu}}{m_{\tau}}\, (G_1 (y_{h^0})+G_1 (y_{A^0})) \Big{\}}  
\,\, ,
\label{tauANOM1}
\end{eqnarray}
and 
\begin{eqnarray}
a_{\mu}^{(2)}(x) &=& \frac{G_F}{\sqrt{2}}\, \frac{Q_{\mu}}{64\,\pi^2}
\,(x-1)^2 \Big{\{} \frac{(\bar{\xi}^{E\,*}_{N,\mu\mu})^2 + 
(\bar{\xi}^{E}_{N,\mu \mu})^2 + 
2\, |\bar{\xi}^{E}_{N,\mu\mu}|^2 \, x}{ 1+(r_{h^0}-2)\,x+x^2}
\nonumber \\ &-&
\frac{(\bar{\xi}^{E\,*}_{N,\mu\mu})^2 + 
(\bar{\xi}^{E}_{N,\mu \mu})^2 - 
2\, |\bar{\xi}^{E}_{N,\mu\mu}|^2\, x}{1+(r_{A^0}-2)\,x+x^2} \Big{\}}
\,\, ,
\label{tauANOM2}
\end{eqnarray}
where $F_1 (w)$ and $G_1 (w)$ are 
\begin{eqnarray}
F_1 (w)&=&\frac{w\,(3-4\,w+w^2+2\,ln\,w)}{(-1+w)^3}\nonumber \,\, , \\
G_1 (w)&=&\frac{w\,(2+3\,w-6\,w^2+w^3+ 6\,w\,ln\,w)}{(-1+w)^4}
\label{functions1}
\end{eqnarray}
Here  $y_{H}=\frac{m^2_{\tau}}{m^2_{H}}$ and 
$r_{H}=\frac{m^2_{H}}{m^2_{\mu}}$, $Q_{\tau}$ and $Q_{\mu}$ are the charges 
of $\tau$ and $\mu$ leptons respectively. In eqs. (\ref{tauANOM1}) and 
(\ref{tauANOM2}) $\bar{\xi}^{E}_{N,ij}$ is defined as 
$\xi^{E}_{N,ij}=\sqrt{\frac{4\,G_F}{\sqrt{2}}}\, \bar{\xi}^{E}_{N,ij}$.
In eq. (\ref{tauANOM}) we take into account internal $\tau$ and
$\mu$-lepton contributions since, the Yukawa couplings 
$\bar{\xi}^{E}_{N, ij}$ $i$( or $j$)$=e$ are negligible (see Discussion 
part). Notice that we make our calculations for arbitrary $q^2$ and take 
$q^2=0$ at the end.   

In our analysis we take the couplings $\bar{\xi}^{E}_{N,\tau\mu}$ and 
$\bar{\xi}^{E}_{N,\mu\mu}$ complex in general and use the parametrization 
\begin{eqnarray}
\bar{\xi}^{E}_{N,l l^{\prime}}=|\bar{\xi}^{E}_{N,l l^{\prime}}|\, 
e^{i \theta_{l l^{\prime}}} \,\, . 
\label{xi}
\end{eqnarray}
The Yukawa factors in eqs. (\ref{tauANOM1}) and  (\ref{tauANOM2}) can 
be written as 
\begin{eqnarray}
((\bar{\xi}^{E\,*}_{N,l l^{\prime}})^2+(\bar{\xi}^{E}_{N, l^{\prime} l})^2)=
2\,cos\,2\theta_{l l^{\prime}}\, |\bar{\xi}^{E}_{N, l^{\prime} l}|^2
\end{eqnarray}
where $l,l^{\prime}=\mu,\tau$. Here $\theta_{l l^{\prime}}$ are CP violating 
parameters which lead to the existence of the lepton electric dipole moment. 
\section{Discussion}
The new physics contribution to the AMM of the lepton is controlled by the 
Yukawa couplings $\bar{\xi}^E_{N,ij}, i,j=e, \mu, \tau$ in the model III. 
These couplings can be complex in general and they are free parameters of 
the model under consideration. The relevant interaction (see eq. (\ref{AMM1}))  
can be created by the mediation of the neutral Higgs bosons $h^0$ and $A^0$
beyond the SM, with internal leptons $e,\, \mu ,\,\tau$ (Fig. \ref{fig1}). 
However, in our predictions, we assume that the Yukawa couplings 
$\bar{\xi}^{E}_{N,\tau\, e}$ and $\bar{\xi}^{E}_{N,\mu\mu}$ are small 
compared to $\bar{\xi}^{E}_{N,\tau\mu}$ since their strength is
proportional to the masses of the leptons denoted by their indices, similar 
to the Cheng-Sher scenerio \cite{Sher}. Notice that, we also 
assume $\bar{\xi}^{E}_{N,ij}$ as symmetric with respect to the indices 
$i$ and $j$. Therefore, the number of free Yukawa couplings is reduced by
two and one more coupling, namely $\bar{\xi}^{E}_{N,\tau\mu}$ still exists 
as a free parameter. This parameter can be restricted by using the 
experimental result of the $\mu$ EDM \cite{Bailey}
\begin{eqnarray}
d_{\mu} < 10.34\times 10^{-19}\, e-cm \,\, .
\label{muedmex}
\end{eqnarray}
at $95\%$ CL limit and the corresponding theoretical result for the EDM of
the muon in the model III (see \cite{Iltan} for details). Since a non-zero 
EDM can be obtained in the case of complex couplings, there exist a CP 
violating parameter $\theta_{\tau\mu}$ coming from the parametrization eq. 
(\ref{xi}). Using the experimental restriction in eq.(\ref{muedmex}), the 
upper limit of the coupling $\bar{\xi}^{E}_{N,\tau\mu}$ is predicted at the 
order of the magnitude of $10^{3}\, GeV$.

The other possibility to get a constraint for the upper limit of 
$\bar{\xi}^{E}_{N,\tau\mu}$ is to use the experimental result of the muon 
AMM. In this work, we study the new physics effects on the muon AMM and 
predict a more stringent bound for the coupling $\bar{\xi}^{E}_{N,\tau\mu}$, 
with the assumption that the new physics effects are of the order of the 
experimental uncertainty of the muon AMM measurement. We also check the 
effect of the coupling $\bar{\xi}^{E}_{N,\mu\mu}$ on AMM of the muon and 
observe that AMM has a weak sensitivity on this coupling. This insensitivity 
is due to the suppression coming from the factors $r_{h^0}$ and $r_{A^0}$ in 
the denominator of eq. (\ref{tauANOM2}). Therefore, we can take  
$\bar{\xi}^{E}_{N,\tau\mu}$ as the only free parameter.
  
\begin{figure}[htb]
\vskip -3.0truein
\centering
\epsfxsize=6.8in
\leavevmode\epsffile{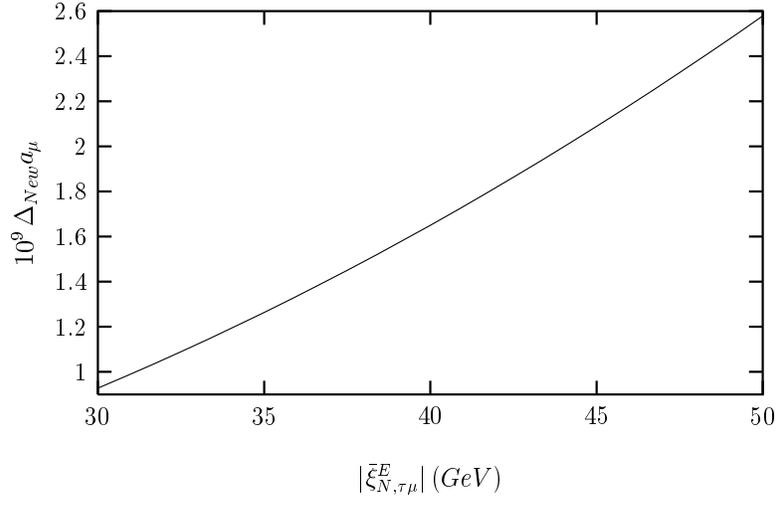}
\vskip -3.0truein
\caption[]{$\Delta_{New} a_{\mu}$ 
as a function of $|\bar{\xi}^{E}_{N,\tau\mu}|$ for $sin\theta_{\tau\mu}=0.5$, 
$m_{h^0}=85\, GeV$ and  $m_{A^0}=95\, GeV$.} 
\label{anommagtauksi}
\end{figure}

Fig. \ref{anommagtauksi} shows the $|\bar{\xi}^{E}_{N,\tau\mu}|$ 
dependence of $\Delta_{New} a_{\mu}$ for $sin\theta_{\tau\mu}=0.5$, 
$m_{h^0}=85\, GeV$ and  $m_{A^0}=95\, GeV$. Here, $\Delta_{New} a_{\mu}$ is
of the order of magnitude $10^{-9}$, increases with increasing
value of the coupling $|\bar{\xi}^{E}_{N,\tau\mu}|$ and exceeds the 
experimental uncertainty, namely $10^{-9}$. This forces us to restrict the 
coupling $|\bar{\xi}^{E}_{N,\tau\mu}|$ as 
$|\bar{\xi}^{E}_{N,\tau\mu}| < 30 \pm 5 \, GeV$ for intermediate values 
of $sin\theta_{\tau\mu}$, $0.4\leq sin\theta_{\tau\mu} \leq 0.6$. This is
a much better upper limit compared to the one obtained using the experimental 
result of the $\mu$ EDM.

\begin{figure}[htb]
\vskip -3.0truein
\centering
\epsfxsize=6.8in
\leavevmode\epsffile{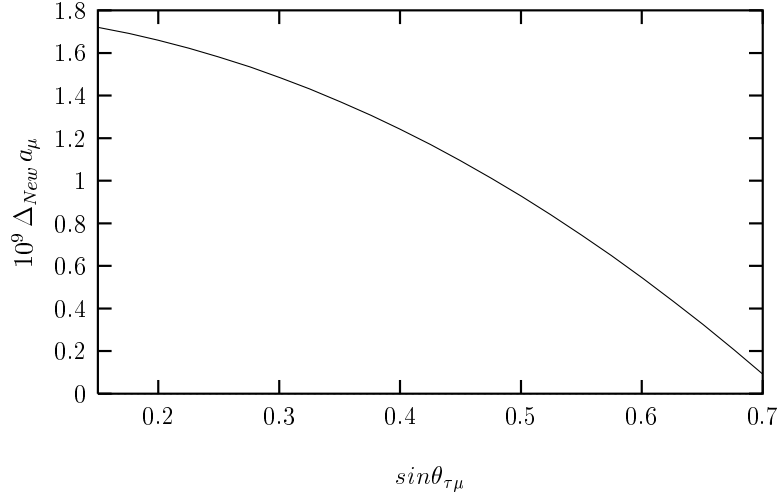}
\vskip -3.0truein
\caption[]{$\Delta_{New} a_{\mu}$ as a function of $sin\,\theta_{\tau\mu}$ 
for $|\bar{\xi}^{E}_{N,\tau\mu}|=30\,GeV$, $m_{h^0}=85\, GeV$ and  
$m_{A^0}=95\, GeV$.} 
\label{anomagtausin}
\end{figure}

In Fig. \ref{anomagtausin}, we show the $sin\theta_{\tau\mu}$ dependence 
of $\Delta_{New} a_{\mu}$ for $\bar{\xi}^{E}_{N,\tau\mu}=30\, GeV$, 
$m_{h^0}=85\, GeV$ and  $m_{A^0}=95\, GeV$. Increasing values of 
$sin\theta_{\tau\mu}$ cause  $\Delta_{New} a_{\mu}$ to decrease and to lie 
within the experimental uncertainty.

\begin{figure}[htb]
\vskip -3.0truein
\centering
\epsfxsize=6.8in
\leavevmode\epsffile{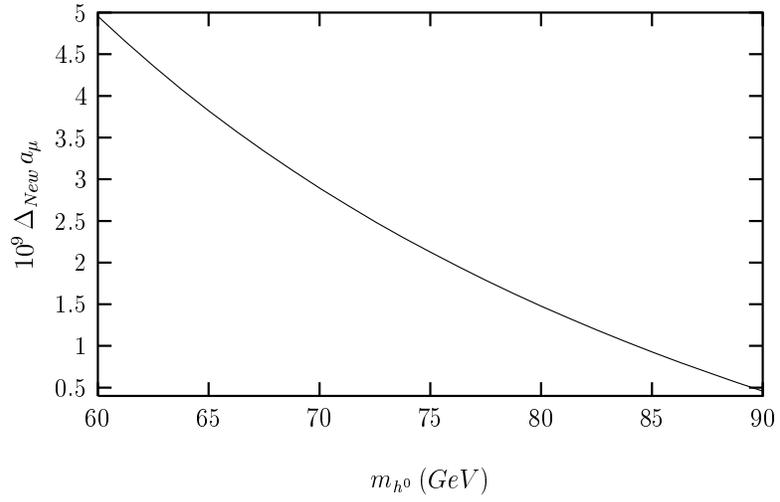}
\vskip -3.0truein
\caption[]{$\Delta_{New} a_{\mu}$ as a function of $m_{h^0}$  for 
$|\bar{\xi}^{E}_{N,\tau\mu}|=30\,GeV$, $m_{A^0}=95\, GeV$ and 
$sin\,\theta_{\tau\mu}=0.5$. .}
\label{anommagh0}
\end{figure}

In Fig. \ref{anommagh0}, we present $m_{h^{0}}$ dependence of 
$\Delta_{New} a_{\mu}$ for $\bar{\xi}^{E}_{N,\tau\mu}=30\, GeV$, 
$sin\theta_{\tau\mu}=0.5$, and  $m_{A^0}=95\, GeV$.
The upper limit of $\Delta_{New} a_{\mu}$ decreases with increasing values 
of $m_{h^0}$. 

For completeness, we also show the $|\xi^{E}_{N,\mu\mu}|$ dependence of
$\Delta_{New} a_{\mu}$ when the internal $\mu$-lepton contribution is taken 
into account. In this figure, it is observed that 
$\Delta_{New} a_{\mu}$ is only weakly sensitive to $|\xi^{E}_{N,\mu\mu}|$ for 
$|\xi^{E}_{N,\mu\mu}| < 0.1 \,GeV$  and therefore the internal $\mu$-lepton 
contribution can be safely neglected, for these values. 

In this work, we choose the type III 2HDM of \cite{Nie} for the physics 
beyond the SM and assume that only FCNC interactions exist at tree level, 
with complex Yukawa couplings. We predict an upper limit for the coupling 
$|\bar{\xi}^{E}_{N,\tau\mu}|$ for the intermediate values of the imaginary 
part, by assuming that the new physics effects are of the order of the 
experimental uncertainty of muon AMM, namely $10^{-9}$, and see that this
leads to a much  better upper limit, $\sim 30 \, GeV$, compared to the one 
obtained by using the experimental result of the $\mu$ EDM, $\sim 10^3 GeV$. 
In the calculations, we studied the internal $\mu$ lepton contributions as
well. However, we observe that they give a negligible contribution to the 
AMM of muon. Furthermore, we neglect the $e$ lepton contribution.  With 
more accurate future measurements of the AMM, it should be possible to
constrain the parameters of the Two Higgs Doublet Model more stringently.
\section{Acknowledgement} 
This work has been supported by the Turkish Academy of Sciences, in the 
framework of the Young Scientist Award Program. (EOI-TUBA-GEBIP/2001-1-8).

\newpage
\newpage
\begin{figure}[htb]
\vskip -3.0truein
\centering
\epsfxsize=6.8in
\leavevmode\epsffile{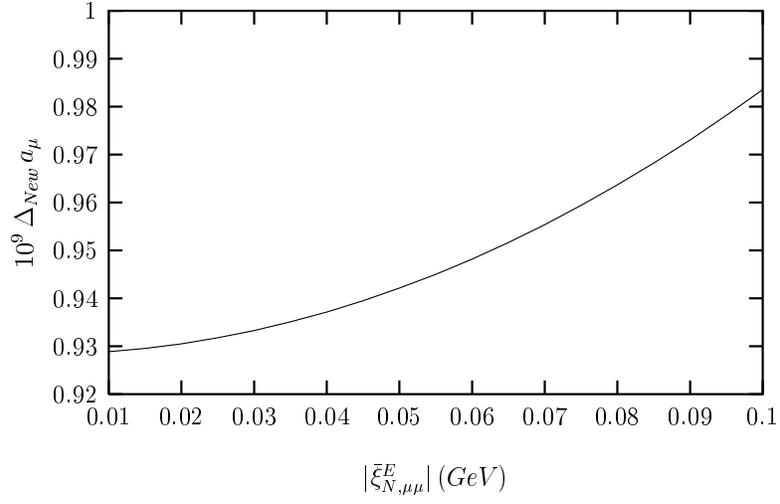}
\vskip -3.0truein
\caption[]{$\Delta_{New} a_{\mu}$ as a function of $|\xi^{E}_{N,\mu\mu}|$ 
for $|\bar{\xi}^{E}_{N,\tau\mu}|=30\,GeV$, $sin\,\theta_{\tau\mu}=0.5$, 
$sin\,\theta_{\mu\mu}=0.5$, $m_{h^0}=85\, GeV$ and  $m_{A^0}=95\, GeV$.}
\label{anommagtotksimiumu}
\end{figure}
\end{document}